\documentclass[11pt]{article}

\usepackage{graphicx}
\usepackage{natbib}
\usepackage{url}

\newtheorem{theorem}{Theorem}
\newtheorem{proof}{Proof}

\newenvironment{pf}{\begin{proof}\normalfont}{\end{proof}}
\newtheorem{remark}{Remark}
\newenvironment{rem}{\begin{remark}\normalfont}{\end{remark}}
\newtheorem{const}{Construction}
\newenvironment{constn}{\begin{const}\normalfont}{\end{const}}

\title{Multi-part balanced incomplete-block designs}
\author{R. A.  Bailey \and Peter J.  Cameron}

\begin{document}

\maketitle

\begin{abstract}
  We consider designs for cancer trials which allow each medical centre
  to treat only a limited number of cancer types with only a limited number
  of drugs.  We specify desirable properties of these designs, and prove
  some consequences.  Then we give several different constructions.  Finally
  we generalize this to three or more factors, such as biomarkers.
\end{abstract}

\section{First design problem}
\subsection{The problem}
This problem was posed by Valerii Fedorov at the workshop on 
\emph{Design and Analysis of Experiments in Healthcare} held at the Isaac Newton
Institute for Mathematical Sciences at Cambridge, U.K.\ in July 2015.
The context is \emph{basket trials}, where several different drugs are treated
on several different diseases in a single protocol which involves many medical
centres: see \cite{orphan} and \cite{Wood}. The combinatorial properties listed
below have been proposed by \cite{VFSL} as potentially giving
optimal designs, which may give a benchmark for designs which are achievable
in practice.

A trial is being designed to compare several drugs for their effects on several
different types of cancer.  In order to keep the protocol simple for each
medical centre involved,
it is proposed to limit each medical centre to only a few of the cancer types
and only a few of the drugs.  Let $v_1$ be the number of cancer types,
$v_2$  the number of drugs, 
and $b$ the number of medical centres. The following properties are desirable:
\begin{enumerate}\itemsep0pt
\item[(a)]
all medical centres involve the same number, say $k_1$, of cancer types,
  where $k_1<v_1$;
\item[(b)]
all medical centres use the same number, say $k_2$, of drugs, where $k_2<v_2$;
\item[(c)]
each pair of distinct cancer types are involved together at the same non-zero 
number, say $\lambda_{11}$, of medical centres;
\item[(d)]
each pair of distinct drugs are used together at the same non-zero 
number, say $\lambda_{22}$, of medical centres;
\item[(e)]
each drug is used on each type of cancer at the same number,
say $\lambda_{12}$,  of medical centres.
\end{enumerate}

The inequalities in conditions~(a) and~(b) force the medical centres to be
incomplete both for cancer types and for drugs.
Insisting that the parameters in conditions (c) and (d) are non-zero
is necessary to prevent the confounding
of either cancer types or drugs with medical centres.

\begin{figure}
\begin{center}
\includegraphics[width=5in]{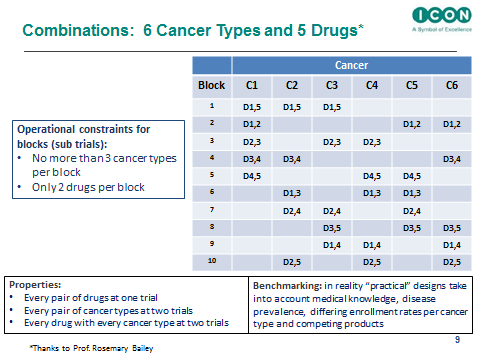}
\end{center}
\caption{Design for 6 cancer types and 5 drugs, using 10 blocks;
each block has 3 cancer types and 2 drugs.}
\label{f1}
\end{figure}

For brevity, from now on the medical centres will be referred to as blocks.
Figure~\ref{f1} shows such a design for six cancer types and five drugs 
using 10 blocks;
it has $k_1=3$ and $k_2=2$.

Conditions~(a) and~(c) specify that the design for cancer types is a 
balanced incomplete-block design, also known as a $2$-design, or,
more specifically, a $2$-$(v_1,k_1,\lambda_{11})$ design.
Likewise, conditions~(b) and~(d)
specify that the design for drugs is a $2$-design.
We call these the C-design and the D-design respectively.

We shall call a design satisfying conditions~(a)--(e) a 
\textit{$2$-part $2$-design} or
\textit{$2$-part balanced incomplete block design}.
These are not the same as the bipartite designs defined by \cite{hoff}.

\subsection{Previous work}
\label{sec:BOMA}
In Section~\ref{sec:2con} we concentrate on designs with only two different 
factors (cancer types and drugs), before generalizing to three or more factors
in Section~\ref{sec2}.  This is partly to help the reader to become familiar 
with the ideas, and partly because this case seems likely to be of practical importance 
in the clinical context described.

The more general case has already been considered by \cite{sitt}, \cite{muk}
and \citet[Section 10.8]{OAbook}.
Because conditions (a)--(d) specify balanced incomplete-block designs and condition~(e)
is reminiscent of the definition of orthogonal multi-array given by \cite{brick}, 
\cite{sitt} called these designs \emph{balanced orthogonal multi-arrays}.
Brickell's original definition was essentially a generalization of
orthogonal arrays of strength two and minimal size, so it included the
conditions that $b$ is a square and $\lambda_{12}=1$.  \cite{sitt} acknowledged
that he was removing those conditions.

However, the original definition of orthogonal multi-array continues to be
in use in many areas.  They give an alternative
definition of semi-Latin squares: see \cite{rabSLS} and \cite{LHS1,LHS2}.
Dually, they are used in factorial designs: see \cite{rabJSTP}.  \cite{WP}
used them in the study of tournaments.  They are used in
cryptography: see, for example, \cite{AMSW} and \cite{seb2}.  Recently,
\cite{mag} have generalized them to strength~$t$, so that $b$ is a $t$-th
power of an integer.  This generalization seems to be within the spirit of the
original definition, whereas Sitter's does not.

Thus we think that ``$2$-part $2$-design'' (or, more generally, a multi-part
$2$-design) is a more suitable name.

\cite{sitt} also  allowed the block size within each type to vary.  \cite{muk}
called the balanced orthogonal multi-arrays \textit{proper} when this is
not allowed.  He also restricted attention to the case where $k_i< v_i$,
unlike \cite{sitt}.  Both allowed $\lambda_{ii}$ to be zero, which permits
confounding: in Table~1 of \cite{muk} one type has its levels confounded with
blocks.

\cite{muk} gave two general constructions for designs of this type. We shall
comment on the relationship of these to our constructions at the
relevant places.

\subsection{Representing the designs}
\label{sec:repn}

How should we represent a design of this type?  Each block has all combinations of
$k_1$ cancer types with $k_2$ drugs, so a full display would show $bk_1k_2$ items.
For example, in the design in Fig.~\ref{f1}, Block 1 contains the ordered pairs
\[
(\mbox{C1}, \mbox{D1}), \quad (\mbox{C1}, \mbox{D5}), \quad
(\mbox{C2}, \mbox{D1}), \quad (\mbox{C2}, \mbox{D5}), \quad
(\mbox{C3}, \mbox{D1}), \quad (\mbox{C3}, \mbox{D5}).
\label{eq:full}
\]
It might be clearer to show these in rectangular form:
\[
\begin{array}{cc}
(\mbox{C1}, \mbox{D1}) & (\mbox{C1}, \mbox{D5}) \\
(\mbox{C2}, \mbox{D1}) & (\mbox{C2}, \mbox{D5})\\
(\mbox{C3}, \mbox{D1}) & (\mbox{C3}, \mbox{D5})
\end{array}
\]
The people running the clinical trial need this full representation.

A dual way to represent the design is to use a $v_1 \times v_2$ rectangle with
$\lambda_{12}$ entries per cell. Equation~(\ref{eqT3}) below shows that this
contains the same number of items as the full representation.
The rows are labelled by the cancer types, and the columns
by the drugs.  The name of each block is shown in each cell $(i,j)$ for which
the combination of cancer type~$i$ and drug~$j$ occurs in that
block.  
Figure~\ref{both}(a) shows the design in Fig.~\ref{f1} in this format. This dual
representation does not extend easily to the generalizations of the problem in 
Sections~\ref{sec2}--\ref{sec:last}.

\begin{figure}
\begin{center}
\begin{tabular}{c@{\qquad}c}
  \begin{tabular}{c||c|c|c|c|c|}
    & D1 & D2 & D3 & D4 & D5\\
    \hline
    \hline
C1 & 1, 2 & 2,  \hphantom{0}3 & 3,  4 & 4, 5 & 1, \hphantom{0}5\\
    \hline
C2 & 1,  6 & 7, 10 & 4, 6 & 4, 7 & 1, 10\\
    \hline
C3 & 1, 9 & 3, \hphantom{0}7 & 3, 8 & 7, 9 & 1, \hphantom{0}8\\
    \hline
C4 & 6, 9 & 3, 10 & 3, 6 & 5, 9 & 5, 10\\
    \hline
C5 & 2, 6 & 2, \hphantom{0}7 & 6, 8 & 5, 7 & 5, \hphantom{0}8\\
    \hline
C6 & 2, 9 & 2, 10 & 4, 8 & 4, 9 & 8, 10\\
    \hline
    \end{tabular}
&
\begin{tabular}{ccc}
Block & Cancer types & Drugs\\
\hline
1 & C1,  C2, C3 & D1, D5\\
2 & C1, C5, C6 & D1, D2\\
3 & C1, C3, C4 & D2, D3\\
4 & C1, C2, C6 & D3, D4\\
5 & C1, C4, C5 & D4, D5\\
6 & C2, C4, C5 & D1, D3\\
7 & C2, C3, C5 & D2, D4\\
8 & C3, C5, C6 & D3, D5\\
9 & C3, C4, C6 & D1, D4\\
10 & C2, C4, C6 & D2, D5
\end{tabular}
\\
\mbox{}\\[-3\jot]
(a) Dual representation & (b) Concise representation
\end{tabular}
\end{center}
\caption{Alternative representations of the design in Fig.~\ref{f1}.
 In the dual representation,
the rows and columns of the rectangle are labelled by cancer types and drugs 
 respectively, and each entry in each cell is the name of a
 block.}
\label{both}
\end{figure}

The most concise way to represent the design is simply to list, for each block, the cancer
types and drugs allocated to it.  This list has $b(k_1+k_2)$ items.
This representation was used by \cite{sitt} and \cite{muk}.
Figure~\ref{both}(b)
gives the concise representation of the design in Fig.~\ref{f1}.

We shall use the concise representation for the remainder of this paper.
However, it can be misinterpreted when removed from the practical context. 
For example, the reader might think that Block~1 in Fig.~\ref{both}(b)
contains five treatments,
those in the union of the sets $\{\mbox{C1},\mbox{C2},\mbox{C3}\}$
and $\{\mbox{D1},\mbox{D5}\}$, rather than
the six treatment combinations in the cartesian product of these sets.
This misinterpretation gives a block design for $v_1+v_2$ treatments in $b$ blocks of
size $k_1+k_2$, which we call the zipped form of the original design.

Figure~\ref{f1} avoids this problem, but at the cost of repeating the
information about the drugs in each block.  This format contains
$bk_1k_2$ items, as many as the full representation, but it seems easier to
read.

Let $N_1$ be the $v_1 \times b$ incidence matrix of cancer types in blocks
in the zipped version of the design.  The $(i,j)$-entry is $1$ if cancer
type~$i$ occurs in block~$j$; otherwise, it is~$0$. Let $N_2$ be the analogous
$v_2 \times b$ incidence matrix for drugs in blocks.  Then the incidence
matrices for the full design are $k_2N_1$ and $k_1N_2$ respectively, not
allowing for the unknown number of times that each combination will
eventually be used in any block.

\subsection{Comparison with other designs}
\label{sec:hum}
At first sight, the design in Fig.~\ref{f1} appears to be a block design
for two treatment factors $C$ and $D$.  However, there are important
differences between this and previous designs.  In our application, the
medical centre represented by Block~$1$ will accept into the trial only
patients with cancer types $1$, $2$ or~$3$.  It has no control in advance
over how many such patients will present themselves.  For each of these
three cancer types, it will randomize approximately equal numbers of
patients to drugs~$1$ and $5$ and placebo, or, in a variant of the original proposal,
approximately one quarter each to placebo, drug~$1$, drug~$5$ and their combination.

\cite{sitt} introduced his designs for use in sampling.  In designed
experiments, \cite{muk} envisaged a completely different sort of application
from the one we describe here.  In that, each block represents a single
observational unit.  For each factor, subsets of the levels are applied, rather
than single levels.  For example, a group of $k_1$ people might be
needed, all playing similar roles, or a hybrid variety of wheat might
be bred from $k_2$ pure lines.  See also \cite{rabSLS}. In this context, it is
not problematic to have $\lambda_{ii}=0$ (so that $k_i=1$) for either
$i=1$ or $i=2$.

In classical factorial designs with blocks of size~$k$,
from \cite{fy}, \cite{rafbook,raf} and \cite{bosefact}
onwards, combinations
of factor  levels do not occur more than once in any block:
thus $k_1=k_2=k$.
Moreover, the subsets
of combinations allocated to blocks are chosen depending on various assumptions
about main effects and interactions.  For example, if $v_1=v_2=3$ and there
are six blocks of three plots each then the design in Fig.~\ref{fact}(a)
permits estimation of both main effects with full efficiency and all
interaction contrasts with efficiency factor $1/2$.

\begin{figure}
\begin{center}
\begin{tabular}{c@{\qquad}c}
    \begin{tabular}{cccc}
      Block 1 & (C1, D1) & (C2, D2) & (C3, D3)\\
      Block 2 & (C1, D2) & (C2, D3) & (C3, D1)\\
      Block 3 & (C1, D3) & (C2, D1) & (C3, D2)\\
      Block 4 & (C1, D1) & (C2, D3) & (C3, D2)\\
      Block 5 & (C1, D2) & (C2, D1) & (C3, D3)\\
      Block 6 & (C1, D3) & (C2, D2) & (C3, D1)
      \end{tabular}
&
    \begin{tabular}{c||c|c|c|}
      & D1 & D2 & D3\\
      \hline
      \hline
      C1 & 1, 4 & 2, 5 & 3, 6\\
      \hline
      C2 & 3, 5 & 1, 6 & 2, 4\\
      \hline
      C3 & 2, 6 & 3, 4 & 1, 5\\
      \hline
      \end{tabular}
\\
\mbox{}\\[-3\jot]
(a) Usual representation & (b) Dual representation
\end{tabular}
\end{center}
  \caption{Classical factorial design for two $3$-level treatment factors
    in $6$~blocks of size~$3$.}
\label{fact}
\end{figure}

The dual form of this design is shown in Fig.~\ref{fact}(b).  The positions
of the block names show clearly how the block design was constructed from a
pair of mutually orthogonal Latin squares. However, unlike in Fig.~\ref{both}(a),
no block name occurs more than once in any row or column.  Consequently,
the occurrences of each block name do not have the rectangular layout that
they do in Fig.~\ref{both}(a).

Later in the twentieth century there was much literature on incomplete-block
designs for two non-interacting treatment factors
with each treatment combination occurring once, so that $v_1v_2=bk$
and $k_1=k_2=k$,
where $k$~is the block size.  For example, \cite{dapbcs}
gave the design in Fig.~\ref{dap}(a).
The dual form is in Fig.~\ref{dap}(b).

\begin{figure}
\begin{center}
\begin{tabular}{c@{\qquad}c}
    \begin{tabular}{lccc}
      Block 1 & (C3, D4) & (C4, D3) & (C6, D1)\\
      Block 2 & (C4, D5) & (C5, D4) & (C6, D2)\\
      Block 3 & (C5, D1) & (C1, D5) & (C6, D3)\\
      Block 4 & (C1, D2) & (C2, D1) & (C6, D4)\\
      Block 5 & (C2, D3) & (C3, D2) & (C6, D5)\\
      Block 6 & (C1, D1) & (C4, D2) & (C3, D5)\\
      Block 7 & (C2, D2) & (C5, D3) & (C4, D1)\\
      Block 8 & (C3, D3) & (C1, D4) & (C5, D2)\\
      Block 9 & (C4, D4) & (C2, D5) & (C1, D3)\\
      Block 10 & (C5, D5) & (C3, D1) & (C2, D4)\\
      \end{tabular}
&
    \begin{tabular}{c||c|c|c|c|c|}
      & D1 & D2 & D3 & D4 & D5\\
      \hline
      \hline
      C1 & \hphantom{0}6 & 4 & 9 & \hphantom{0}8 & \hphantom{0}3\\
      \hline
      C2 & \hphantom{0}4 & 7 & 5 & 10 & \hphantom{0}9\\
      \hline
      C3 & 10 & 5 & 8 & \hphantom{0}1 & \hphantom{0}6\\
      \hline
C4 &  \hphantom{0}7 & 6 & 1 & \hphantom{0}9 & \hphantom{0}2\\
\hline
C5 & \hphantom{0}3 & 8 & 7 & \hphantom{0}2 & 10\\
\hline
C6 & \hphantom{0}1 & 2 & 3 & \hphantom{0}4 & \hphantom{0}5\\
\hline 
      \end{tabular}
\\
\mbox{}\\[-3\jot]
(a) Usual representation &  (b) Dual representation
\end{tabular}
\end{center}
\caption{Block design for two non-interacting sets of treatments, with $v_1=6$, $v_2=5$, 
$b=10$ and $k=3$.}
\label{dap}
\end{figure}

Many authors required the block design for each treatment factor separately
to be balanced.  This is the analogue of conditions~(a)--(d) when $k_1=k_2$.
From \cite{ag:sdr} and \cite{dapbka} onwards, another condition was often imposed,
eventually called \textit{adjusted orthogonality} by \cite{eccruss:77}: the product
$\tilde{N}_1\tilde{N}_2^\top$ should have all its entries equal, where
$\tilde{N}_1$ and $\tilde{N}_2$ are the
$v_1\times b$ and $v_2\times b$ incidence matrices
for the first and second treatment factors, respectively, in blocks.
Although this is a consequence of
condition~(e), it is not equivalent to it.  The duals of designs satisfying
these
conditions were called \textit{triple arrays} by \cite{trip}. The design in
Fig.~\ref{dap}(b) is a triple array.

In spite of the similar conditions that they satisfy,
triple arrays are not special cases of $2$-part $2$-designs, nor vice
versa.
In  a triple array, no block name occurs more than once in
any row or column.
In the dual form of a $2$-part $2$-design, any block name that occurs in
a given row must occur $k_2$ times in that row.
A consequence of the ``non-zero part'' of condition~(d) is  that $k_2>1$.

Apart from the designs given by \cite{DAPWY}, infinite families of triple
arrays have proved frustratingly hard to find: see \citet[Section 13]{bcc}.
By contrast, in Sections~\ref{sec:2con} and~\ref{sec:last}
of this paper we give many simple
constructions of $2$-part $2$-designs and their generalizations.

\subsection{Conditions on parameters}
\label{sec:thm}

An ordinary block design is said to be $\alpha$-resolved if its set of blocks can be partitioned 
into classes in such a way  that each treatment occurs $\alpha$ times in each class.  This
terminology does not extend easily to $2$-part $2$-designs, because cancer types may occur
in different numbers of blocks from drugs. We propose calling a $2$-part block design 
\textit{$c$-partitionable}
if the set of blocks can be grouped into $c$ classes of $b/c$ blocks each,
in such a way that every cancer type occurs the same number of times in each class and
every drug occurs the same number of times in each class.

\begin{theorem}
  \label{th1}
  If there is a $2$-part $2$-design with the parameters given in
  conditions~(a)--(e), then
each cancer type 
occurs in $r_1$ blocks
and each drug
occurs in $r_2$ blocks,
where 
\begin{equation}
r_1=bk_1/v_1, \qquad r_2=bk_2/v_2.
\label{eqT1}
\end{equation}
Moreover, the following equations are satisfied:
\begin{equation}
  v_1(v_1-1)\lambda_{11}=bk_1(k_1-1), \qquad 
  v_2(v_2-1)\lambda_{22}=bk_2(k_2-1),
\label{eqT2}
\end{equation}
and
\begin{equation}
  bk_1k_2=v_1v_2\lambda_{12},
  \label{eqT3}
  \end{equation}
as well as the inequality
\begin{equation}
  b\ge v_1+v_2-1.
  \label{eqT4}
  \end{equation}
If the design is $c$-partitionable then
\begin{equation}
b \geq v_1 + v_2 +c -2.
\label{eq:aha}
\end{equation}
\end{theorem}

\begin{pf}
  The first two statements are the usual conditions for the $2$-designs
 on cancer types and 
drugs respectively, while
Equation~(\ref{eqT3}) equates two different ways of counting the
number of choices of a cancer type, a drug, and a block containing both.

For inequality (\ref{eq:aha}),
let $N=(N_1^\top,N_2^\top,N_0^\top)^\top$,
where $N_1$ and $N_2$ are the incidence matrices defined in
Section~\ref{sec:repn} and $N_0^\top$ is the $b\times c$ incidence matrix of
blocks in classes.
Then
\[NN^\top =
\left[
  \begin{array}{c@{\quad}c@{\quad}c}
    (r_1-\lambda_{11})I+\lambda_{11}J & \lambda_{12}J & (r_1/c)J\\
    \lambda_{12}J & (r_2-\lambda_{22})I+\lambda_{22}J & (r_2/c)J\\
    (r_1/c)J & (r_2/c)J & (b/c)I
    \end{array}
  \right],
\]
where $I$ and $J$ are identity and all-$1$ matrices of the appropriate sizes.

We claim that $NN^\top$ has rank $v_1+v_2+c-2$, from which
inequality~(\ref{eq:aha}) follows.
First, let ${w}_1$, $w_2$ and ${w}_3$ be column vectors of lengths 
$v_1$, $v_2$, $c$ respectively whose entries sum to $0$. Then
\begin{equation}
  NN^\top
  \left(\begin{array}{c}
      w_1\\w_2\\w_3\end{array}
      \right)
  = \left(\begin{array}{c}
      (r_1-\lambda_{11}){w}_1\\
    (r_2-\lambda_{22}){w}_2\\
    (b/c)w_3
    \end{array}
\right).
  \label{eq1}
\end{equation}
Because the blocks are incomplete, $\lambda_{11}< r_1$ and $\lambda_{22} <r_2$,
and
so the restriction of this matrix to the space of such vectors,
which has dimension $v_1+v_2+c-3$, is invertible.
The orthogonal complement of this space consists of all vectors
of the form $(xj_1^\top,yj_2^\top,zj_3^\top)^\top$, where $j_1$, $j_2$ and $j_3$
are all-$1$ vectors of lengths $v_1$, $v_2$ and $c$ respectively.
The action of
$NN^\top$ on this space is obtained by replacing the block matrices by their
row sums: using the results in (\ref{eqT1})--(\ref{eqT3}), this simplifies to
\[
\left(
  \begin{array}{ccc}
    r_1k_1 & r_1k_2 & r_1\\ r_2k_1 & r_2k_2 & r_2\\
    (b/c)k_1 & (b/c)k_2 & (b/c)
    \end{array}
  \right),
\]
which has rank~$1$. So the claim~(\ref{eq:aha}) is proved.

The first part of the theorem shows that every $2$-part $2$-design is $1$-partitionable.
Thus inequality~(\ref{eqT4}) is a special case of inequality~(\ref{eq:aha}).
\end{pf}

\begin{rem}
  \cite{muk} remarked on the integrality conditions (\ref{eqT1})--(\ref{eqT3})
  without stating them explicitly, and proved inequality (\ref{eqT4}).
  \end{rem}

\begin{rem} 
  Inequality~(\ref{eqT4}) can be
  regarded as a generalization of both Fisher's and Bose's inequalities:
  see \citet[Chapter 1]{cvl} and \citet[Chapter 11]{doce}.
For Fisher's inequality, take the C-design to be any $2$-design with $v=v_1$, 
and take a single drug which occurs in all blocks; we have $\lambda_{12}=r_1$ and
$\lambda_{22}=0$: although our
conditions that $\lambda_{22}>0$ and $k_2<v_2$
fail for the D-design, the
proof still works, because the only vector ${w}_2$ in equation~(\ref{eq1})
is the zero vector: thus the proof
gives $b\ge v+1-1$. For Bose's inequality, take the
C-design to be any resolvable $2$-design
with $v=v_1$ and replication $r=r_1$, and the drugs 
to be labelled by the resolution classes of the design, with a drug in every 
block in the corresponding resolution class, so that $v_2=r$.
We have $\lambda_{12}=1$ and $\lambda_{22}=0$.
Again 
part of condition~(d)
fails, but the proof works, 
giving $b\ge v+r-1$.
Inequality~(\ref{eq:aha}) seems to be the true analogue of Bose's inequality
for $2$-part $2$-designs.
\end{rem}

\begin{rem}
  Although neither triple arrays nor $2$-part $2$-designs are special cases
  of the other, they both satisfy inequality~(\ref{eqT4}).
Proofs for triple arrays are in
\cite{Bagchi1998}, \cite{bcc} and \cite{trip},
and the proof that we have given here also works for triple arrays.
\end{rem}

\section{Constructions of $2$-part $2$-designs}
\label{sec:2con}

In this section, we give several constructions. In order to identify when
two different constructions give designs which are essentially the same,
we say that two $2$-part $2$-designs are \textit{isomorphic} to each other
if one
can be obtained from the other by relabelling some of blocks, cancer types
and drugs.  
Weak isomorphism generalizes this by also allowing the roles of cancer types
and drugs to be interchanged.

Given two or more non-isomorphic designs for the same parameters,
there may be practical reasons for preferring one over the rest.

Since interchanging roles does not affect conditions (a)--(e), from now
on we usually adopt the convention that
\begin{equation}
  v_1\geq v_2.
  \label{eq:exchange}
  \end{equation}

Given a $2$-part $2$-design, the procedure of \textit{C-swap} creates a new 
$2$-part $2$-design. This
simply involves replacing the set of cancer treatments 
in each block with the complementary
set. This changes the parameters $k_1$, $\lambda_{11}$ and $\lambda_{12}$ to
$v_1-k_1$, $b-2r_1+\lambda_{11}$ and $r_2-\lambda_{12}$, 
leaving $b$, $v_1$, 
$v_2$, $k_2$ and $\lambda_{22}$ unchanged. The new design fulfills all the 
conditions so long as $v_1-k_1\geq 2$.
The combination of a C-swap and the
analogous D-swap has the effect of replacing each block by its complement
(in the zipped version).

Thus, in our search for design constructions, we may assume that
\begin{equation}
\mbox{for $i=1$ and $i=2$,
either $k_i\leq v_i/2$ or $k_i=v_i-1$.}  
  \label{eq:swap}
  \end{equation}
All of our tables
are limited to parameter sets which
satisfy conditions (\ref{eq:exchange}) and (\ref{eq:swap}).

\begin{constn} \textbf{(Cartesian products)}
\label{c:cart}
One obvious method of construction is the cartesian product. This
starts with two balanced incomplete-block designs,
one for $v_1$ treatments in $b_1$ blocks of size $k_1$, 
the other for $v_2$ treatments in $b_2$ blocks of size $k_2$. 
Form all $b_1b_2$ combinations of a block of each sort.
For each combination, form the cartesian product of their subsets of treatments.
\end{constn}

This will usually result in rather large values of $b$.  For example, when 
$v_1=6$, $k_1=3$, $v_2=5$ and $k_2=2$ then the smallest possible values of 
$b_1$ and $b_2$ are both $10$, so this construction gives a design with 
$100$ blocks, unlike the design with $10$ blocks in Fig.~\ref{f1}.

Table~\ref{t2} shows the parameters of the 
designs with the least number of blocks which can be 
constructed by this method with $v_1\geq v_2$, using the table of $2$-designs in Appendix~I of
\cite{hall};
note that design~13 in that table should have $k=4$.

\begin{table}
\caption{Parameter sets for the designs with the least number of blocks
that can be made by 
Construction 1: 
$v_1$ is the number of cancer types, $v_2$ is the number of drugs, 
and $b$ is the number of blocks, each of which has
$k_1$ cancer types and $k_2$ drugs}
\label{t2}
\[
\addtolength{\arraycolsep}{-0.5\arraycolsep}
\begin{array}{rrrrr}
b & v_1 & v_2 & k_1 & k_2\\
\hline
9 & 3 & 3 & 2 & 2\\
12 & 4 & 3 & 3 & 2\\
15 & 5 & 3 & 4 & 2\\
16 & 4 & 4 & 3 & 3\\
18 & 4 & 3 & 2 & 2\\
18 & 6 & 3 & 5 & 2
\end{array}
\qquad
\begin{array}{rrrrr}
b & v_1 & v_2 & k_1 & k_2\\
\hline
20 & 5 & 4 & 4 & 3\\
21 & 7 & 3 &  3 & 2\\
21 & 7 & 3 & 6 & 2\\\mbox{}
24 & 4 & 4 & 3 & 2\\
24 & 6 & 4 & 5 & 3\\
\mbox{}
\end{array}
\qquad
\begin{array}{rrrrr}
b & v_1 & v_2 & k_1 & k_2\\
\hline
24 & 8 & 3 & 7 & 2\\
25 & 5 & 5 & 4 & 4\\
27 & 9 & 3 & 8 & 2\\
28 & 7 & 4  & 3 &3\\
28 & 7 & 4 & 6 & 3\\
\mbox{}
\end{array}
\qquad
\begin{array}{rrrrr}
b & v_1 & v_2 & k_1 & k_2\\
\hline
30 & 5 & 3 & 2 & 2\\
30 & 5 & 4 & 4 & 2\\
30 & 6 & 3 & 3 & 2\\
30 & 6 & 5 & 5 & 4\\
32 & 8 & 4 & 7 & 3\\
\mbox{}
\end{array}
\]
\end{table}

\begin{constn} \textbf{(Subcartesian products)}
\label{c:subcart}
If  $k_2$ divides $v_2$ then there may exist a resolved $2$-design
$\Delta_2$ for $v_2$ drugs in $b_2$ blocks of size~$k_2$  
with $r$ resolution classes.  Suppose that $\Delta_1$ is a $2$-design
for $v_1$ cancer types in $b_1$ blocks of size~$k_1$,
where $b_1$ is a multiple of~$r$.
Now we can
achieve a $2$-part $2$-design without taking the full product.
Partition the blocks of $\Delta_1$ into $r$ classes of size $b_1/r$
in any way at all, and match these classes
to the resolution classes of $\Delta_2$ in any way.  For each matched pair, construct 
the cartesian product design.  Putting these products together gives a design of the required type 
with $b_1b_2/r$ blocks, considerably fewer than the $b_1b_2$ blocks in the entire 
product of $\Delta_1$ and $\Delta_2$. 

More generally, if the design $\Delta_2$ is $c$-partitionable and $c$ divides
$b_1$ then replace the resolution classes in this construction by the $c$
classes of blocks. This gives a $2$-part $2$-design with $b_1b_2/c$ blocks.
Putting $c=1$ gives Construction~\ref{c:cart} as a special case of this.
\end{constn}

Figures~\ref{f5two}(a) and~(b) show two possibilities when $v_1=v_2=4$,
$k_1=k_2=2$ and $r=3$.

\begin{figure}
\begin{center}
\begin{tabular}{c@{\qquad}c}
\begin{tabular}{ccc}
Block & Cancer types & Drugs\\
\hline
1 & C1,  C2 & D1, D3\\
2 & C1, C2 & D2, D4\\
3 & C3, C4 & D1, D3\\
4 & C3, C4 & D2, D4\\
5 & C1, C3 & D2, D3\\
6 & C1, C3 & D1, D4\\
7 & C2, C4 & D2, D3\\
8 & C2, C4 & D1, D4\\
9 & C1, C4 & D1, D2\\
10 & C1, C4 & D3, D4\\
11 & C2, C3 & D1, D2\\
12 & C2, C3 & D3, D4
\end{tabular}
&
\begin{tabular}{ccc}
Block & Cancer types & Drugs\\
\hline
1 & C1,  C2 & D1, D3\\
2 & C1, C2 & D2, D4\\
3 & C1, C3 & D1, D3\\
4 & C1, C3 & D2, D4\\
5 & C1, C4 & D2, D3\\
6 & C1, C4 & D1, D4\\
7 & C2, C3 & D2, D3\\
8 & C2, C3 & D1, D4\\
9 & C2, C4 & D1, D2\\
10 & C2, C4 & D3, D4\\
11 & C3, C4 & D1, D2\\
12 & C3, C4 & D3, D4
\end{tabular}
\\
\mbox{}\\[-3\jot]
(a)  Constructions 2 and 3 & (b) Construction 2 but not Construction 3 
\end{tabular}
\end{center}
\caption{Two designs for 4 cancer types and 4 drugs, using 12 blocks;
each block has 2 cancer types and 2 drugs.}
\label{f5two}
\end{figure}

Table~\ref{t3} shows some  parameter sets for designs that can be made by
Construction 2, possibly after an interchange or a swap, with $k_i\leq 10$ for
$i=1$ and $i=2$.

There are two special cases.
When $b_1=r$ then we simply match the blocks of $\Delta_1$ to
the resolution classes of $\Delta_2$.
When $v_1=3$, $k_1=2$,  $v_2=4$, $k_2=2$ and
$r=3$, this gives the design in Fig.~\ref{f2}. 
When $v_1=v_2=6$, $k_1=k_2=3$ and $r=10$, this gives the design in
Fig.~\ref{f20}(a).
When $v_1=7$, $k_1=3$, $v_2=15$, $k_2=3$ and $r=7$, this gives a $2$-part
$2$-design with $b=35$, $r_1=5$, $r_2=7$, $\lambda_{11}=5$, $\lambda_{22}=1$
and $\lambda_{12}=3$.

\begin{figure}
\begin{center}
\begin{tabular}{ccc}
Block & Cancer types & Drugs\\
\hline
1 & C1,  C2 & D1, D3\\
2 & C1, C2 & D2, D4\\
3 & C1, C3 & D2, D3\\
4 & C1, C3 & D1, D4\\
5 & C2, C3 & D1, D2\\
6 & C2, C3 & D3, D4
\end{tabular}
\end{center}
\caption{Design for 3 cancer types and 4 drugs, using 6 blocks;
  each block has 2 cancer types and 2 drugs.}
\label{f2}
\end{figure}

On the other hand, 
if $\Delta_1$ is also resolved with replication $r$ then
we may match the resolution classes of the two designs.  For example, 
when $v_1=v_2=4$ and $k_1=k_2=2$ then we may take $r=3$ and $b_1=b_2=6$ 
to get the design in Fig.~\ref{f5two}(a).  This is not even weakly isomorphic to the design
in Fig.~\ref{f5two}(b), where the pairs of blocks from $\Delta_1$ do not form
resolution classes.
When $v_1/k_1 = v_2/k_2=2$ and $b_1=b_2$,
Construction 3 also gives designs with these parameters.

\begin{figure}
  \begin{center}
    \begin{tabular}{@{}cc}
\begin{minipage}[t]{6cm}
  \begin{tabular}{ccc}
Block & Cancer types & Drugs\\
\hline
1 & C1,  C2, C3 & D1, D5, D6\\
2 & C1, C2, C3 & D2, D3, D4\\
3 & C1, C5, C6 & D1, D2, D6\\
4 & C1, C5, C6 & D3, D4, D5\\
5 & C1, C3, C4 & D2, D3, D6\\
6 & C1, C3, C4 & D1, D4, D5\\
7 & C1, C2, C6 & D3, D4, D6\\
8 & C1, C2, C6 & D1, D2, D5\\
9 & C1, C4, C5 & D4, D5, D6\\
10 & C1, C4, C5 & D1, D2, D3\\
11 & C2, C4, C5 & D1, D3, D6\\
12 & C2, C4, C5 & D2, D4, D5\\
13 & C2, C3, C5 & D2, D4, D6\\
14 & C2, C3, C5 & D1, D3, D5\\
15 & C3, C5, C6 & D3, D5, D6\\
16 & C3, C5, C6 & D1, D2, D4\\
17 & C3, C4, C6 & D1, D4, D6\\
18 & C3, C4, C6 & D2, D3, D5\\
19 & C2, C4, C6 & D2, D5, D6\\
20 & C2, C4, C6 & D1, D3, D4
\end{tabular}
\end{minipage}
&
\begin{minipage}[t]{8cm}
  \begin{tabular}{cccc}
Block & Cancer types & Drugs & Biomarkers\\
\hline
1 & C1,  C3, C5 & D1, D4, D5& B1, B2\\
2 & C2, C4, C6 & D2, D3, D6 & B1, B2\\
3 & C1, C5, C6 & D3, D5, D6 & B1, B3\\
4 & C2, C3, C4 & D1, D2, D4 & B1, B3\\
5 & C1, C2, C3 & D2, D3, D5 & B1, B4\\
6 & C4, C5, C6 & D1, D4, D6 & B1, B4\\
7 & C1, C4, C5 & D1, D2, D3 & B1, B5\\
8& C2, C3, C6 & D4, D5, D6 & B1,  B5\\
9& C1, C4, C6 & D2, D4, D5 & B2, B3\\
10& C2, C3, C5 & D1, D3, D6 &  B2, B3\\
11 & C1, C3, C4 & D3, D4, D6 & B2, B4\\
12& C2, C5, C6 & D1, D2, D5& B2, B4\\
13 & C1, C2, C5 & D2, D4, D6 & B2, B5\\
14& C3, C4, C6 & D1, D3, D5 & B2, B5\\
15& C1, C2, C4 & D1, D5, D6 & B3, B4\\
16& C3, C5, C6 & D2, D3, D4 & B3, B4\\
17 & C1, C2, C6 & D1, D3, D4 & B3, B5\\
18 & C3, C4, C5 & D2, D5, D6 & B3, B5\\
19& C1, C3, C6 & D1, D2, D6 & B4, B5\\
20 & C2, C4, C5 & D3, D4, D5 & B4, B5
  \end{tabular}
\end{minipage}\\
(a) Construction 2 & (b) Construction 3 (ignoring the $5$ biomarkers)\\
&
followed by Construction 9\\
    \end{tabular}
    \caption{Two designs for 6 cancer types and  6 drugs,
      using 20 blocks.
    \label{f20}}
    \end{center}
\end{figure}

At first sight,
the two general constructions given by \cite{muk} are special cases
of this. His first construction needs both $\Delta_1$ and $\Delta_2$ to be
$c$-partitionable, and matches the classes.
This includes the cartesian product when $c=1$,
and when $c=3$ it gives the design in Figure~\ref{f5two}(a) but not the one
in Figure~\ref{f5two}(b).  His second construction use a $c$-partitionable
design $\Delta_2$ only when $b_1=c$.  However, if $c$ divides $b_1$ then we
may replace $\Delta_2$ by $b_1/c$ copies of it, giving a $b_1$-partionable
design whose classes can be matched to the blocks of $\Delta_1$.

Thus Construction~\ref{c:subcart} is precisely equivalent to
the combination of the two in \cite{muk}.

\begin{constn} \textbf{(Hadamard matrices)}
\label{c:had}
Some
$2$-part $2$-designs in which $v_1=v_2$ and $k_1=k_2 =v_1/2$
arise from Construction~2. Another way of getting such designs is to 
start with a Hadamard matrix $H$ of order $4n$, where $n = k_1$, 
in which the elements in the first row are all $+1$. Identify the $2n$ cancer 
types with the columns in which the second row has
entry $+1$, and identify the $2n$ drugs with the columns in which the second 
row has entry $-1$.  Each of the remaining rows gives two blocks, one 
containing all the objects whose columns have entries $+1$, and one containing 
all the objects whose columns have entries $-1$.  Thus $b=8n-4$.
Moreover, each pair of blocks contains each cancer type and each drug just once,
in the concise representation, so the $2$-part $2$-design is 
$(4n-2)$-partitionable and the lower bound in inequality~(\ref{eq:aha}) is achieved.
\end{constn}

For example, when $n=3$ we can take
\[
H = \left[
\begin{array}{cccccccccccc}
+1 & +1 &+1 &+1 &+1 &+1 &+1 &+1 &+1 &+1 &+1 &+1\\
+1 &+1 &+1 &+1 &+1 &+1 &-1 & -1 & -1 & -1 & -1 & -1\\
+1 & -1 & +1 & -1 & +1 & -1 & +1 & -1 & -1 & +1 & +1 & -1\\
+1 & -1 & -1 & -1 & +1 & +1 & -1 & -1 & +1 & -1 & +1 & +1\\
+1 & +1 & +1 & -1 & -1 & -1 & -1 & +1 & +1 & -1 & +1 & -1\\
+1 & -1 & -1 & +1 & +1 & -1 & +1 & +1 & +1 & -1 & -1 & -1\\
+1 & -1 & -1 & +1 & -1 & +1 & -1 & +1 & -1 & +1 & +1 & -1\\
+1 & -1 & +1 & +1 & -1 & -1 & -1 & -1 & +1 & +1 & -1 & +1\\
+1 & +1 & -1 & -1 & +1 & -1 & -1 & +1 & -1 & +1 & -1 & +1\\
+1 & +1 & -1 & +1 & -1 & -1 & +1 & -1 & -1 & -1 & +1 & +1\\
+1 & +1 & -1 & -1 & -1 & +1 & +1 & -1 & +1 & +1 & -1 & -1\\
+1 & -1 & +1 & -1 & -1 & +1 & +1 & +1 & -1 & -1 & -1 & +1
\end{array}
\right].
\]
Labelling the columns as $C_1$, \ldots, $C_6$, $D_1$, \ldots, $D_6$ in order, the
construction gives the design in
the first three columns of Fig.~\ref{f20}(b), ignoring the biomarkers.
It is not weakly isomorphic to the design in Fig.~\ref{f20}(a),
because all triples
of cancer types and all triples of drugs occur.

\begin{table}
\caption{Parameter sets for the designs with the least number of blocks
that can be made by 
Constructions \ref{c:subcart} or \ref{c:had} but not \ref{c:cart}:
$v_1$ is the number of cancer types, 
$v_2$ is the number of drugs, and $b$ is the number of blocks, each of which has
$k_1$ cancer types and $k_2$ drugs; $r$ is a number used in the construction.
Asterisks denote the only parameter sets for designs  achievable by Construction~3}
\label{t3}
\[
\addtolength{\arraycolsep}{-0.7\arraycolsep}
\begin{array}{crrrrrr}
 & b & v_1 & v_2 & k_1 & k_2 & r\\
\hline
 & 6 & 4 & 3 & 2 & 2 & 3\\
{*} & 12 & 4 & 4 & 2 & 2 & 3\\
 & 12 & 9 & 4 & 3 & 3 & 4\\
 & 14 & 8 & 7 & 4 & 3 & 7\\
 & 14 & 8 & 7 & 4 & 6 & 7\\
 & 15 & 6 & 5 & 2 & 4 & 5\\
& 18 & 9 & 4 & 8 & 2 & 3\\
 & 20 & 6 & 5 & 3 & 2 & 10\\
 {*} & 20 & 6 & 6 & 3 & 3 & 5\\
 & 20 & 16 & 5 & 4 & 4 & 5\\
 & 22 & 12 & 11 & 6 & 5 & 11\\
 & 22 & 12 & 11 & 6 & 10 & 11\\
& 24 & 9 & 4 & 3 & 2 & 3\\
& 24 & 9 & 8 & 3 & 7 & 4\\
\end{array}
\quad
\begin{array}{crrrrrr}
& b & v_1 & v_2 & k_1 & k_2 & r\\
\hline
 & 28 & 8 & 7 & 2 & 3 & \hphantom{0}7\\
 & 28 & 8 & 7 & 2 & 6 & 7\\
{*} & 28 & 8 & 8 & 4 & 4 & 7\\
& 30 & 6 & 4 & 2 & 2 & 3\\
& 30 & 6 & 5 & 2 & 2  & 5\\
& 30 & 6 & 6 & 3 & 2 & 5\\
& 30 & 10 & 4 & 4 & 2 & 3\\
& 30 & 15 & 4 & 7 &  2 & 3\\ 
 & 35 & 15 & 7 & 3 & 3 & 7\\
 & 35 & 15 & 7 & 3 & 6 & 7\\
& 36 & 9 & 4 & 4 & 2 & 3\\
& 36 & 9 & 9 & 3 & 3 & 4\\
& 36 & 10 & 4 & 9 & 2 & 3\\
{*} & 36 & 10 & 10 & 5 & 5 & 9
\end{array}
\quad
\begin{array}{crrrrrr}
& b & v_1 & v_2 & k_1 & k_2 & r\\
\hline
& 40 & 16 & 5 & 4 & 2 & 5\\
& 40 & 16 &  6 & 4 & 3& 5\\
& 42 & 7 & 4 & 2 & 2 & 3\\
& 42 & 8 & 7 & 4 & 2 & 7\\
& 42 & 21 & 4 & 5 & 2 & 3\\
& 42 & 21 & 8 & 5 & 4 & 7\\
{*} & 44 & 12 & 12 & 6 & 6& 11\\
& 45 & 6 & 6 & 2 & 2 & 5\\
& 45 & 10 & 6 & 4 & 2 & 5\\
& 45 & 15 & 6 & 7 & 2 & 5\\
& 48 & 16 & 4 & 6 & 2 & 3\\
& 48 & 16 & 9 & 6 & 3 & 4\\
\ast & 52 & 14 & 14 & 7 & 7& 13\\
& 54 & 27 & 4 &  13  & 2 & 3\\
\end{array}
\quad
\begin{array}{crrrrrr}
& b & v_1 & v_2 & k_1 & k_2 & r\\
\hline
& 56 & 8 & 8 & 4 & 2 & 7\\
& 60 & 10 & 4 & 3 & 2 & 3\\
& 60 & 10 & 6 & 3 & 3 & 10\\
& 60 & 16 & 4 & 8 & 2 & 3\\
& 60 & 16 & 6 & 4 & 2 & 5\\
& 60 & 16 & 6 & 8 & 3 & 10\\
& 60 & 16 & 9  & 4 & 3 & 4\\
& 60 & 16 & 10 & 4 & 4 & 5\\
& 60 & 16 & 15 & 4 & 7 & 5\\
{*} & 60 & 16 & 16 & 8 & 8& 15\\
& 60 & 21 & 4 & 7 & 2 & 3\\
& 60 & 21 & 6 & 7 & 3 & 10\\
& 60 & 25 & 4 & 5 & 2 & 3\\
& 60 & 25 & 6 & 5 & 3 & 10\\
\end{array}
\]
\end{table}

The asterisked entries in
Table~\ref{t3} show the parameters of the smallest designs that can be
constructed by this method.

When $n=4$ this construction gives the design in
Fig.~\ref{f5two}(a).  For some values of $n$, different choices of Hadamard
matrix, or different designations of which row is second, can give
non-isomorphic designs.  It may be that there are some values of $n$ for which there
exists a Hadamard matrix of order $4n$ but no $2$-$(2n,n,n-1)$ design.  If so,
Construction~3 gives a design for these parameters but Construction~2 does not.
Such a value of $n$ is likely to be too large to affect designs of practical size.
However, we retain this construction, because it produces resolvable designs,
which can be used as ingredients in Construction~\ref{c:mprod} in
Section~\ref{sec2} to give designs without too many blocks.

\begin{constn} \textbf{(Symmetric $2$-designs)}
Here is another  general method of construction. Consider  a symmetric balanced
incomplete-block design $\Delta$ for $v$ treatments in $v$ blocks of size $k$.
Every pair
of distinct treatments concur in $\lambda$ blocks, where $\lambda = k(k-1)/(v-1)$,
and every pair of distinct blocks have $\lambda$ treatments in common.  Let $\Gamma$
be one block of $\Delta$.  Identify the treatments in $\Gamma $ with $k$ drugs $D_1$,
\ldots, $D_k$ and the remaining treatments with $v-k$ cancer types $C_1$, \ldots,
$C_{v-k}$.  Now consider the design~$\Delta'$ consisting of all blocks of $\Delta$ except
$\Gamma$.  Each of these blocks contains $\lambda$ drugs and $k-\lambda$ 
cancer types.  In $\Delta'$, each pair of drugs concur in $\lambda-1$ blocks; each pair
of cancer types concur in $\lambda$ blocks; and each drug occurs with each cancer
type in $\lambda$ blocks. 
Thus $b = v-1$, $v_1=v-k$, $v_2=k$, $k_1=k-\lambda$, $k_2=\lambda$, 
$\lambda_{11} = \lambda_{12}= \lambda$ and $\lambda_{22} = \lambda -1$.
\end{constn}

We can use Construction~4 whenever there exists a symmetric $2$-$(v,k,\lambda)$ design
with $v=v_1+v_2$, $k=v_2$ and $\lambda=k_2$, provided that $k_1+k_2=v_2$.
In order to satisfy condition~(d),
$\lambda$~must be bigger than one.
The lower bound in inequality~(\ref{eqT4}) is always met.

The properties of symmetric $2$-designs guarantee that conditions~(c) and (d)
hold, but they also match up the blocks of the C-design and the D-design,
which typically produces fewer blocks than previous construction methods.

The design in Fig.~\ref{f1} can be obtained by this construction with $v=11$, $k=5$
and $\lambda=2$.   Figure~\ref{f2} gives the design with $v=7$, $k=4$ and $\lambda=2$.

Table~\ref{t1} lists parameter sets for small designs that can be constructed
by this method, with an interchange and swaps where necessary:
again using Table~I.1 in \cite{hall}. After allowing for possible interchanges
and swaps, this table represents $34$ designs.

\begin{table}
\caption{Parameter sets for which small designs can be made by Construction~4: 
$v_1$ is the number of cancer types, $v_2$ is the number of drugs, and $b$ 
is the number of blocks, each of which has
$k_1$ cancer types and $k_2$ drugs;  $v$, $k$ and $\lambda$ are parameters of the 
symmetric $2$-design used in the construction}
\label{t1}
\[
\addtolength{\arraycolsep}{-0.8\arraycolsep}
\begin{array}{rrrrrrrr}
b & v_1 & v_2 & k_1 & k_2 &v & k & \lambda\\
\hline
6 & 4 & 3 & 2 & 2 & 7 & 4 & 2\\
10 &6 & 5& 3 & 2 & 11 & 5 & 2\\
12 & 9 & 4 & 6 & 3 & 13 & 9 & 6\\
\end{array}
\qquad
\begin{array}{rrrrrrrr}
b & v_1 & v_2 & k_1 & k_2 &v & k & \lambda\\
\hline
14 &8 & 7 & 4 & 3 &15 & 7 & 3\\
15 &10 & 6 & 4 & 2 & 16 & 6 & 2\\
18 & 10 & 9 & 5 & 4 & 19 & 9 & 4\\
\end{array}
\qquad
\begin{array}{rrrrrrrr}
b & v_1 & v_2 & k_1 & k_2 &v & k & \lambda\\
\hline
22 & 12 & 11 & 6 & 5 & 23 & 11 & 5\\
24 & 16 & 9 & 6 & 3 & 25 & 9 & 3\\
\mbox{}
\end{array}
\]
\end{table}

\begin{constn} \textbf{(Augmentation)}
Given a $2$-part $2$-design~$\Delta$ in which $v_2=2k_2+1$,
we may augment it to one
for one more drug by increasing $v_2$ to $v_2+1$ and $k_2$ to $k_2+1$ while
merely doubling the number of blocks.  Replace each block of $\Delta$ by
two blocks, both with the same set of cancer types as before.  One of these
blocks has the previous set of drugs and the extra drug, while the other has
all the remaining drugs.
\end{constn}

For example, augmenting the design in Fig.~\ref{f1}
gives the design in Fig.~\ref{f20}(a).

Applying the augmentation just to the D-design gives a resolvable $3$-design,
as shown in the Extension Theorem of \cite{alltop}. This can be used directly
in Construction~\ref{c:subcart}. However, augmentation is such a straightforward
way of obtaining one $2$-part $2$-design from another that we think it is 
worth identifying.

\begin{constn} \textbf{(Group-divisible designs)}
\label{c:gd}
If $v_1=v_2=v$ and $k_1=k_2=k$ then the zipped form of
a $2$-part $2$-design is a semi-regular group-divisible incomplete-block design
for two so-called groups of $v$ treatments in blocks of size $2k$ with $k>1$:
see \cite{srgd}.
Unzipping any one of these gives a $2$-part $2$-design.
\end{constn}

Table VII of \cite{clat} gives three such designs. Unzipping them gives
the product design for the first parameter set in Table~\ref{t2},
the design in Fig.~\ref{f5two}(a) and the design in the first three columns of
Fig.~\ref{f20}(b).

\begin{constn} \textbf{(Group actions)}
\label{c:ga}
Here is a construction based on group actions. Suppose that the group $G$ acts
$2$-transitively on two sets $C$ and $D$ of sizes $v_1$ and $v_2$ respectively, and that $G$ is also transitive in the induced action on $C\times D$. 
Choose a subset of $C$ and a
subset of $D$, each containing at least two points; their union is a block,
and the images of this block under $G$ give the remaining blocks. 
The blocks have to be unzipped to give a $2$-part $2$-design.
This does
not give much control over $b$, except that we know it is a divisor of the
order of $G$. A strategy for finding good designs by this method is to choose
a subgroup $H$ of $G$ which acts intransitively on each of $C$ and $D$, and
to use fixed sets of $H$ on $C$ and $D$ in the construction. 
\end{constn}

The three examples below arise from this construction, but can be more easily
be derived from the $3$-$(22,6,1)$ design $\Xi$ whose automorphism
group is the Mathieu group $M_{22}$. 
It has $22$ points and $77$ blocks of size $6$, any two
blocks meeting in zero or two points;
see \citet[Chapters 1 and 9]{cvl}.
For simplicity, we describe the cancer types as red points and the
drugs as green points.

Take a block $B_0$ of the design $\Xi$; its points are red, 
and the remaining $16$
points are green. For each of the $60$ blocks meeting $B_0$ in two points,
we define a block of our new design containing two red and four green points.
Now two red points lie in five blocks, one of which is $B_0$; so they lie
in four more blocks.
A red and green point lie in five blocks, each containing two red points.
Two green points lie in three blocks meeting $B_0$. For each point of
$B_0$ lies in a unique such block, and each block contains two points of $B_0$.
So we have an example with $v_1=6$, $v_2=16$, $b=60$, $k_1=2$, $k_2=4$, and
$(\lambda_{11},\lambda_{12},\lambda_{22})=(4,5,3)$.

Table \ref{t3} shows these parameters for a design made by Construction 2
and an interchange.

The other two examples use the $4$-$(23,7,1)$ design $\Theta$ in which  the
blocks through the extra point are formed by adjoining that point to the
blocks of $\Xi$: see \cite{cvl}. The counting arguments that verify their
properties are similar to what we have just seen.

For the second design, we take a set $A$ of seven points which form
a block of $\Theta$ not containing the extra point. These will be red, 
and the remaining $15$ points of $\Xi$ green. Any block of $\Xi$ meets $A$ in
one or three points; we take the blocks meeting $A$ in three points to be the
blocks of the required design.
We obtain an example with $v_1=7$, $v_2=15$, $b=35$, $k_1=k_2=3$, and
$(\lambda_{11},\lambda_{12},\lambda_{22})=(5,3,1)$.

This has the same parameters as the
fifth design made using Construction~\ref{c:subcart}.

Finally, using the $23$-point design $\Theta$ but not throwing away the extra
point we obtain a design with $v_1=7$, $v_2=16$, $k_1=3$, $k_2=4$,
$(\lambda_{11},\lambda_{12},\lambda_{22})=(20,15,7)$, $b=140$. Another design
with these parameters is the cartesian product of the projective plane of
order 2 and the affine plane of order 4; these designs are not isomorphic.

To build these from the group action construction, the relevant groups are the
stabilizers of the sets of six or seven red points in the appropriate Mathieu
groups; these are the groups $2^4\colon S_6$, $A_7$, and $2^4\colon A_7$
respectively. 

\section{Generalizing the design problem}
\label{sec2}

\subsection{The extended problem}
In March 2016 Valerii Fedorov extended the problem as follows.  Can we add a 
third factor,
whose levels are biomarkers in this case, subject to the obvious extra 
conditions?  Here we generalize this to an arbitrary number $m$ of factors.

The conditions for an $m$-part $2$-design are as follows.  The analogue
of conditions (a)--(b) is that, for $1\leq i \leq m$,
factor~$i$ has $v_i$ levels
and each medical centre involves $k_i$ of them,
where $k_i<v_i$; the analogue of conditions (c)--(d) is that, for
$1\leq i\leq m$, each pair of
levels of factor~$i$ are used together at the
same non-zero number $\lambda_{ii}$ of medical centres.

The generalization of condition~(e) is less clear.  When $i=3$,
a weak generalization is that each biomarker is used on each cancer type
at the same number $\lambda_{13}$ of medical centres and  that each biomarker 
is used with each drug at the same number $\lambda_{23}$ of medical centres. 
For now, we use this weak version. Note, however, that this gets us into the 
territory of factorial design, so we might be
confounding all or part of a two-factor interaction with all or part of a main effect.
By analogy with orthogonal arrays \citep{OAbook}, we call this weak generalization a
\textit{$3$-part $2$-design with strength~$2$}, whereas a
$3$-part $2$-design with strength~$3$ would have every triple of 
(cancer type, drug, biomarker) at the same number $\lambda_{123}$ of medical 
centres.

Thus the strength-$2$ generalization of condition~(e) is that, for $1 \leq i <j
\leq m$, each
level of factor~$i$ occurs with each level of factor~$j$
at the same number $\lambda_{ij}$ of medical centres.

\subsection{Conditions on parameters in the extended problem}

\begin{theorem}
  In an $m$-part $2$-design of strength~$2$, all of the following are satisfied.
  \begin{itemize}
    \item[(i)]
  The analogues of equations~(\ref{eqT1}) and~(\ref{eqT2}) hold for 
  each factor.
  \item[(ii)]
Equation~(\ref{eqT3}) generalizes to
$bk_ik_j = v_iv_j\lambda_{ij}$ for $1\leq i<j\leq m$.
\item[(iii)]
  If the design is $c$-partitionable then $b \geq v_1+ \cdots + v_m +c-m$.
  \item[(iv)]
    In particular, $b \geq v_1 + \cdots + v_m - m+1$.
    \end{itemize}
\end{theorem}

Proofs are similar to those in Section~\ref{sec:thm}.
Part~(iv) is precisely Theorem~1 of \cite{muk}.

\section{Constructions of $m$-part $2$-designs of strength at least~$2$}
\label{sec:last}

\subsection{Two main constructions}

Here we give the main construction of \cite{muk} in the language of this paper.

\begin{constn} \textbf{(Orthogonal arrays)}
\label{c:oa}
Suppose that there is a positive integer~$c$ such that, for $i=1$, \ldots, $m$,
$\Delta_i$ is a $c$-partitionable $2$-design for $v_i$ treatments in
$b_i$ blocks of size~$k_i$.  Moreover, there is an orthogonal array $\Gamma$
with $m$ columns, where column~$i$ contains $b_i/c$ symbols for
$1\leq i \leq m$.

Match the $c$ classes of blocks of $\Delta_1$, \ldots, $\Delta_m$.
For $j=1$, \ldots, $c$ separately, each row $\rho$ of $\Gamma$ gives a block
of the new design, as follows.  For $i=1$, \ldots, $m$,
identify the block in class~$j$ of
$\Delta_i$ labelled by the symbol in row $\rho$ and column~$i$ of $\Gamma$:
then form the cartesian product of these $m$ blocks.
This gives a $c$-partitionable $m$-part $2$-design in $sc$ blocks,
where $s$ is the number of rows of $\Gamma$.  The strength of this new
design is equal to the strength of the orthogonal array $\Gamma$.

In one extreme case, $\Gamma$ has all possible different rows, so that
$s=\prod_{i=1}^m b_i/c^m$.  If, in addition, $c=1$, then $s=\prod_{i=1}^m b_i$
and we obtain the full cartesian product.
\end{constn}

The design in Fig.~\ref{f8and9}(a) can be made in this way with $c=1$,
using an orthogonal array with three columns, each with three symbols.

An example with $m=3$ and $c=2$ is shown in  Fig.~\ref{f8and9}(b),
which is contained in Table~1 of \cite{muk}.
Here, $v_i=4$, $k_i=2$ and $b_i=6$ for $i=1$, $2$, $3$,
and each of $\Delta_1$, $\Delta_2$ and $\Delta_3$ can be resolved into
pairs of blocks. For each design, label the replicates $1$, $2$, $3$
in any order.
For $j=1$, $2$, $3$, combine the $j$-th replicates from the three designs, not
by the full cartesian product, which would give eight blocks, but by using an
orthogonal array of strength~$2$ with four rows and three columns,
each with two symbols.
This $3$-part $2$-design has strength $2$ but not strength $3$.

\begin{figure}
  \begin{center}
\begin{tabular}{@{}cc}
  \begin{minipage}[t]{6cm}
    \begin{tabular}{cccc}
      & Cancer & & Bio-\\
Block &  types & Drugs & markers\\
\hline
1 & C1, C2 & D1, D2 & B1, B2\\
2 & C1, C3 & D1, D3 & B1, B3\\
3 & C2, C3 & D2, D3 & B2, B3\\
4 & C1, C2 & D2, D3 & B1, B3\\
5 & C1, C3 & D1, D2 & B2, B3\\
6 & C2, C3 & D1, D3 & B1, B2\\
7 & C1, C2 & D1, D3 & B2, B3\\
8 & C1, C3 & D2, D3 & B1, B2\\
9 & C2, C3 & D1, D2 & B1, B3\\
\mbox{}\\
\mbox{}\\
\mbox{}
    \end{tabular}
  \end{minipage}
  &
  \begin{minipage}[t]{6cm}
        \begin{tabular}{cccc}
      & Cancer & & Bio-\\
Block &  types & Drugs & markers\\
\hline
1 & C1, C2 & D1, D2 & B1, B2\\
2 & C1, C2 & D3, D4 & B3, B4\\
3 & C3, C4 & D1, D2 & B3, B4\\
4 & C3, C4 & D3, D4& B1, B2\\
5 & C1, C3 & D1, D3 & B1, B3\\
6 & C1, C3 & D2, D4 & B2, B4\\
7 & C2, C4 & D1, D3 & B2, B4\\
8 & C2, C4 & D2, D4 & B1, B3\\
9 & C1, C4 & D1, D4 & B1, B4\\
10 & C1, C4 & D2, D3 & B2, B3\\
11 & C2, C3 & D1, D4 & B2, B3\\
12 & C2, C3 & D2, D3 & B1, B4
\end{tabular}
  \end{minipage}
  \\
  (a) $3$ cancer types, $3$ drugs, $3$ biomarkers
  &
    (b) $4$ cancer types, $4$ drugs, $4$ biomarkers
    \end{tabular}
    \end{center}
  \caption{Two $3$-part $2$-designs, both made from Construction~8}
  \label{f8and9}
\end{figure}

Table~1 of \cite{sitt} gives a $7$-part $2$-design made in this way with
$b=24$ and $v_i=2k_i=4$ for $i=1$, \ldots, $7$.

\begin{constn} \textbf{(Products of multi-part designs)}
\label{c:mprod}
The  ingredients of the previous construction are $m$ individual $2$-designs
and an orthogonal array, which may be trivial.
Instead, we may start with multi-part $2$-designs, or an assortment of
$2$-designs and multi-part $2$-designs. The use of orthogonal arrays and/or
$c$-partitioning can be extended to this method too. As in
Construction~\ref{c:subcart}, we can allow one of the constituent designs to be
not $c$-partitionable, as long as its number $b$ of blocks is divisible by $c$.
\end{constn}

The full product of
an $m_1$-part $2$-design $\Theta_1$ with $b_1$ blocks
and an $m_2$-part $2$-design $\Theta_2$ with $b_2$ blocks is an $(m_1+m_2)$-part
$2$-design with $b_1b_2$ blocks and strength~$2$.  If $\Theta_1$ has strength
$m_1$ and $m_2=1$ then the full product has strength $m_1+1$.
For example, if  $m=3$, $v_3=3$ and $k_3=2$ then the product of the
design in Fig.~\ref{f1} and a $2$-design with 
three blocks of size $2$ gives a $3$-part $2$-design with $30$ blocks and
strength~$3$.

As an example of the relaxation of the $c$-partitionable condition, suppose that
$\Theta$ is a $c$-partitionable $m_1$-part $2$-design for drugs and cancer
types and $\Delta$ is a $2$-design for $v_3$ biomarkers in $c$ blocks of size
$k_3$. We can simply match the blocks of $\Delta$ to the classes of $\Theta$
in any way. The $3$-part $2$-design in Fig.~\ref{f20}(b) was made like this by
starting with a $2$-part $2$-design made by Construction~\ref{c:had}, grouping
blocks into ten classes of the form $\{2i-1,2i\}$, and matching these classes 
to the ten blocks of a $2$-design $\Delta$ for five biomarkers.

The special case of this with $b=c$ is the second general construction
given by \cite{muk}. As noted in Section~\ref{sec:2con}, this specialization
does not restrict his  designs.  However, because we have now given more
constructions for the case that $m=2$, applying the various product
constructions to them produces new designs for higher values of $m$ also.

\subsection{Other constructions}

The augmentation method in Construction~5
easily generalizes to three or more factors.
If $v_i =2k_i +1$ then $v_i$ and $k_i$ can be increased by one
while the number of blocks is merely doubled.

If $v_1= \cdots =v_m = v$ and $k_1= \cdots =k_m=k$ then the zipped form
of an $m$-part $2$-design is a semi-regular group-divisible design for $mv$
treatments in blocks of size $mk$ with $k>1$.
Just as in Construction~6,
any such design can be unzipped to give a $m$-part $2$ design.
There are two such designs with $m=3$
in Table~VII of \cite{clat}. Their unzipped forms
are the designs in Fig.~\ref{f8and9}.
The one with $m=4$ gives the design in Figure~\ref{fig:new}, which can also be
obtained from a $9\times 4$ orthogonal array with three symbols in each column.

\begin{figure}
\begin{center}
\begin{tabular}{ccccc}
& Cancer & & Bio- \\
Block & types & Drugs & markers & Activity\\
\hline
1 & C1, C2 & D1, D2 & B1, B2 & A1, A2\\
2 & C1, C2 & D2, D3 & B2, B3 & A1, A3\\
3 & C1, C2 & D1, D3 & B1, B3 & A2, A3\\
4 & C1, C3 & D1, D2 & B2, B3 & A2, A3\\
5 & C1, C3 & D2, D3 & B1, B3 & A1, A2\\
6 & C1, C3 & D1, D3 & B1, B2 & A1, A3\\
7 & C2, C3 & D1, D2 & B1, B3 & A1, A3\\
8 & C2, C3 & D2, D3 & B1, B2 & A2, A3\\
9 & C2, C3 & D1, D3 & B2, B3 & A1, A2\\
\end{tabular}
\end{center}
\caption{A $4$-part $2$-design}
\label{fig:new}
\end{figure}

The group method in Construction 7 also easily extends to
three or more factors:
simply take a permutation group with more than two
$2$-transitive actions.

\section*{Acknowledgement}
We thank Valerii Fedorov for posing this interesting
problem and for the illustration in Fig.~\ref{f1}.

\end{document}